\documentclass[preprint,a4paper]{elsarticle}

\PassOptionsToPackage{hyphens}{url}\usepackage[hyperindex,
linktocpage,
bookmarksnumbered,
plainpages=false,
pdfpagelabels,
colorlinks,
bookmarks,
linkcolor=black,
urlcolor=black,
citecolor=black,
pdfpagemode=UseOutlines,
hyperfootnotes = false,
breaklinks=true]{hyperref}

\usepackage[labelfont=bf,justification=raggedright,singlelinecheck=false]{caption}
\usepackage{float}
\usepackage{lineno,hyperref}
\usepackage{subfig}
\usepackage{color}
\usepackage{amsfonts}
\renewcommand{\vec}[1]{\ensuremath{\boldsymbol{#1}}} 
\graphicspath{{figures/}}
\usepackage[cmex10]{amsmath}
\modulolinenumbers[5]

\journal{TBD}

\usepackage{framed} 
\usepackage{multicol} 

\usepackage{nomencl} 
\makenomenclature
\renewcommand*\nompreamble{\begin{multicols}{2}}
\renewcommand*\nompostamble{\end{multicols}}

\usepackage{soul}

\begin{document}

\begin{frontmatter}

\title{Reduced and Aggregated Distribution Grid Representations Approximated by Polyhedral Sets}

\author[fen]{P. Fortenbacher \corref{cor1}}
\ead{fortenbacher@fen.ethz.ch}
\author[fen]{T. Demiray}
\ead{demirayt@fen.ethz.ch}

\cortext[cor1]{Corresponding author}

\address[fen]{Research Center for Energy Networks (FEN), ETH Zurich, Sonneggstrasse 28, 8092 Zurich }

\begin{abstract}
In this paper we present a novel tractable method to compute reduced and aggregated distribution grid representations that provide an interface in the form of active and reactive power (PQ) capability areas for improving transmission service operator -- distribution service operator (TSO--DSO) interactions. Based on a lossless linear power flow approximation we derive polyhedral sets to determine a reduced PQ operating region capturing all voltage magnitude and branch power flow constraints of the entire distribution grid. To demonstrate the usefulness of our method, we compare the capability area obtained from the polyhedral approximation with an area generated by multiple optimal power flow (OPF) solutions for different distribution grids. While the approximation errors are reasonable, especially for low voltage (LV) grids, the computational complexity to compute the PQ capability area can be significantly reduced with our proposed method.
\end{abstract}


\end{frontmatter}

	
	\nomenclature{$\Delta\theta_\mathrm{d}$}{angular difference design parameter}
	\nomenclature{$\Delta\vec{\theta},\Delta\vec{v}$}{decision vector voltage angles and magnitudes}
	\nomenclature{$\Delta v_\mathrm{d}$}{magnitude difference design parameter}
	\nomenclature{$\epsilon_\theta, \epsilon_v$}{RMS errors on voltage angles and magnitudes}
	\nomenclature{$\epsilon_{\Delta\theta}, \epsilon_{\Delta v}$}{RMS errors on voltage angle and magnitude differences}
	\nomenclature{$\theta_k \in \vec{\theta} $}{decision vector voltage angles}
	\nomenclature{$\theta_\mathrm{s}$}{transformer shift angle}
	\nomenclature{$\vec{a}_q \in \vec{A}_q$}{polygon shape parameters}
	\nomenclature{$b \in \vec{b} $}{per unit line susceptance vector}
	\nomenclature{$\vec{b}_\theta,\vec{b}_v$}{binary decision vector associated with $\Delta\vec{\theta},\Delta\vec{v}$}
	\nomenclature{$b_\mathrm{c}$}{charge susceptance}
	\nomenclature{$\vec{C}_\mathrm{br}$}{branch-node incidence matrix}
	\nomenclature{$\vec{C}_\mathrm{f}$}{branch-from incidence matrix}
	\nomenclature{$\vec{C}_\mathrm{g}$}{generator incidence matrix}
	\nomenclature{$\vec{C}_\mathrm{t}$}{branch-to incidence matrix}
	\nomenclature{$f_p,f_q$}{linear or quadratic costfunction on active and reactive generator power}
	\nomenclature{$f_z$}{objective value of problem $z$}
	\nomenclature{$g \in \vec{g}$}{per unit line conductance vector}
	\nomenclature{$k_1,k_2$}{linear loss gradient parameters}
	\nomenclature{$M$}{big M constant}
	\nomenclature{$n_\mathrm{b},n_\mathrm{g},n_\mathrm{l}$}{number of buses, generators, and lines}
	\nomenclature{$p_k \in \vec{p},\vec{p}_\mathrm{d} $}{nodal active power and load}
	\nomenclature{$\vec{p}_\mathrm{f}$}{active power flow at from ends}
	\nomenclature{$\vec{p}_\mathrm{g} $}{decision vector active generator power}
	\nomenclature{$p_{\ell}$}{active power loss}
	\nomenclature{$\vec{p}_\ell^{\Delta\theta},\vec{p}_\ell^{\Delta v}$}{active power loss vectors generated by voltage angle differences and magnitudes}
	\nomenclature{$q_k \in \vec{q}, \vec{q}_\mathrm{d}$}{nodal reactive power and load}
	\nomenclature{$\vec{q}_\mathrm{f}$}{reactive power flow at from ends}
	\nomenclature{$\vec{q}_\mathrm{g} $}{decision vector reactive generator power}
	\nomenclature{$q_{\ell}$}{reactive power loss}
	\nomenclature{$\vec{q}_\ell^{\Delta\theta},\vec{q}_\ell^{\Delta v}$}{reactive power loss vectors generated by voltage angle differences and magnitudes}
	\nomenclature{$r$}{per unit tap ratio}
	\nomenclature{$\vec{s}$}{apparent power line limit}
	\nomenclature{$v_k \in \vec{v}$}{decision vector voltage magnitudes}
	\nomenclature{$\vec{x},\vec{x}',\vec{x}''$}{decision vectors LOLIN-OPF, LIN-OPF, MIP-OPF}
	\nomenclature{$\vec{x}_\mathrm{pf},\vec{x}_\mathrm{s}$}{solution vectors from power flow and optimization}
	\nomenclature{$y$}{complex series admittance}
	\nomenclature{$y_{ff}^i,y_{ft}^i, y_{tf}^i,y_{tt}^i$}{admittance parameters to describe standard $\pi$ branch model}
	\nomenclature{$y_{ff}'^i,y_{tt}'^i$}{adjusted admittance parameters}
	\nomenclature{$\vec{Y}_\mathrm{b}$}{nodal admittance matrix}
	\nomenclature{$\vec{Y}'_\mathrm{b}$}{adjusted nodal admittance matrix}
	\nomenclature{$\vec{Y}_\mathrm{f},\vec{Y}_\mathrm{t}$}{branch-from and -to admittance matrices}
	\nomenclature{$\vec{Y}'_\mathrm{f},\vec{Y}'_\mathrm{t}$}{adjusted branch-from and -to admittance matrices}
	\nomenclature{$y_\mathrm{sh}$}{shunt admittance}
	
	\printnomenclature


\section{Introduction}
According to the ENTSO-E \cite{ENTSO-E2015} TSOs and DSOs should establish a market for consumers to participate in ancillary services. This also includes voltage services. Nowadays, to enable a secure grid operation, the TSO determines a day-ahead voltage schedule for transmission nodes. Based on this schedule the participating generators need to adjust and provide their reactive power at delivery time and will be remunerated by the TSO \cite{swissgrid} in Switzerland. DSOs can enter an active or a semi-active role as described in \cite{swissgrid}. In these roles they get penalized if they do not operate their grid within a certain active (P) and reactive (Q) power tolerance band. Since in future also distribution grid units might contribute to provide voltage support for upper grid levels, there is a need to have a PQ capability area of the distribution grid describing the available flexibility for the TSO. Such area  (managed and calculated by the DSO) could inform the TSO to which extent the participating units can provide reactive power in the same way as large generators entering an active voltage support role. The advantage of such reduced and aggregated distribution grid representations is three-fold. First, we can reduce the complexity for the TSO, since instead of all distribution grid constraints only a few constraints need to be considered by the TSO. Secondly, the data exchange between TSO and DSO is reduced and thirdly data privacy is ensured.    

\subsection{Related work and Contribution}
The idea to represent a distribution grid as a PQ capability area is not novel, i.e. the author of \cite{Capitanescu2018} forms such capability regions by solving several optimal power flow (OPF) problems at given supporting points. In contrast to this, we developed a method that does not require the search of supporting points and the solutions of multiple OPF problems. Instead, we use linear power flow approximations \cite{Fortenbacher2019} to compute a polytope that defines the PQ capability area, which reduces the computational complexity significantly. As another advantage, our formulation could be directly incorporated as linear constraints in OPF problems to represent distribution grids. In this way, it is also possible to represent distribution grids directly in market clearing problems. The contribution of this paper is two-fold. First, we derive a more detailed formulation of the standard Power Transfer Distribution Factors (PTDFs) that also captures the impact of nodal reactive powers. Second, we derive the polyhedral sets to compute the polytope for the PQ capability areas.

\section{Method}
We mainly use the linear power and branch flow approximation presented in~\cite{Fortenbacher2019}. Here, we summarize the main results, for full details see \cite{Fortenbacher2019}. As an extension, we derive transfer distribution factors (TDFs) for voltage magnitudes (VTDF), angles ($\theta$TDF) and active /reactive power branch flows (PTDF) as a function of the nodal active and reactive power.   

\subsection{Linear Power Flow Approximation}
As derived in \cite{Fortenbacher2019} the lossless power flow approximation in a network with $n_b$ buses and $n_l$ branches is
\begin{equation}
\left[\begin{array}{cc} -\Im\{\vec{Y}_\mathrm{b}'\} & \Re\{\vec{Y}_\mathrm{b} \} \\ -\Re\{\vec{Y}_\mathrm{b}'\} & -\Im\{\vec{Y}_\mathrm{b}\}\ \end{array} \right] \left[\begin{array}{c}\vec{\theta} \\ \vec{v} \end{array} \right]  = 
\left[ \begin{array}{c} \vec{p} \\ \vec{q} \end{array} \right] \ ,
\label{eq:linapprox}
\end{equation} 
\noindent where $\vec{Y}_\mathrm{b} \in \mathbb{C}^{n_b \times n_b}$ is the nodal admittance matrix and $\vec{Y}_\mathrm{b}' \in \mathbb{C}^{n_b \times n_b}$ its adjusted version to correctly represent the power flow equations. $\vec{p},\vec{q} \in \mathbb{R}^{n_b\times 1}$ are the nodal and reactive powers, $\vec{v},\vec{\theta} \in \mathbb{R}^{n_b\times 1}$ are the voltage magnitudes and angles. 

To describe the impact of the powers on the voltage magnitudes and angles, we need to invert \eqref{eq:linapprox} as follows: 
\begin{align}
\left[\begin{array}{c}\vec{\theta} \\ \vec{v} \end{array} \right]  = \underbrace{\left[\begin{array}{cc} -\Im\{\vec{Y}_\mathrm{b}'^{(\theta_0,v_0)}\} & \Re\{\vec{Y}_\mathrm{b}^{(\theta_0,v_0)} \} \\ -\Re\{\vec{Y}_\mathrm{b}'^{(\theta_0,v_0)}\} & -\Im\{\vec{Y}_\mathrm{b}^{(\theta_0,v_0)}\}\ \end{array} \right]^{-1}}_{:= \left[\begin{array}{cc} {\Theta TDF} \in \mathbb{R}^{n_b-1 \times 2(n_b-1)} \\ {VTDF} \in \mathbb{R}^{n_b-1 \times 2(n_b-1)} \end{array} \right] }\left[ \begin{array}{c} \vec{p} \\ \vec{q} \end{array} \right] 	+ \left[\begin{array}{c} \vec{1} \theta_0 \\ \vec{1} v_0\end{array} \right]	.\label{eq:vapprox}	  
\end{align}
Since the full admittance matrix is in general rank-deficient, we need to delete the corresponding columns and rows associated with the slack bus meaning $\vec{Y}_\mathrm{b}^{(\theta_0,v_0)},\vec{Y}_\mathrm{b}'^{(\theta_0,v_0)} \in \mathbb{C}^{n_b-1 \times n_b-1}$, and $\vec{p},\vec{q},\vec{\theta},\vec{v} \in \mathbb{R}^{n_b-1 \times 1}$. To recover the original values of $\vec{\theta}$ and $\vec{v}$, we need to add the deleted slack bus voltage magnitude and angle $v_0,\theta_0 \in \mathbb{R}^{1 \times 1}$ to them.

\subsection{Linear Branch Flow Approximation}
In accordance with \cite{Fortenbacher2019} and with the $\theta$TDF and VTDF matrix, we can express the branch flow active and reactive powers $\vec{p}_\mathrm{f},\vec{q}_\mathrm{f} \in \mathbb{R}^{n_l \times 1}$ as  
\begin{equation}
\left[\begin{array}{c}\vec{p}_\mathrm{f} \\ \vec{q}_\mathrm{f} \end{array} \right] = \underbrace{\left[\begin{array}{cc} -\Im\{\vec{Y}_\mathrm{f}'\} & \Re\{\vec{Y}_\mathrm{f} \} \\ -\Re\{\vec{Y}_\mathrm{f}'\} & -\Im\{\vec{Y}_\mathrm{f}\}\ \end{array} \right]\left[\begin{array}{cc} {\Theta TDF} \\ {VTDF} \end{array} \right]}_{:=PTDF \in \mathbb{R}^{2n_l \times 2(n_b-1)}} \left[ \begin{array}{c} \vec{p} \\ \vec{q} \end{array} \right] , \label{eq:branchapprox}
\end{equation}
where  $\vec{Y}_\mathrm{f},\vec{Y}_\mathrm{f}' \in \mathbb{C}^{n_l \times n_b-1}$ are the slack bus adjusted branch-from admittance matrices. Note that this extended PTDF matrix is a more detailed version of the standard PTDF matrix, since it also captures power flows influenced by the nodal reactive powers.  

\subsection{PQ Capability Area Mapping}
The interface between the distribution and the transmission grid is the distribution feeder injection point connected to a transmission node. The nodal distribution powers $\vec{p},\vec{q}$ are mapped to the controllable aggregated feeder power $P, Q$ as follows
\begin{equation}
\left[ \begin{array}{c} \vec{p} \\ \vec{q} \end{array} \right] =  \underbrace{\left[\begin{array}{cc}\vec{C}_\mathrm{g}  GSK_p & \vec{0} \\ \vec{0} &   \vec{C}_\mathrm{g}GSK_q \end{array} \right]}_{:=\vec{T}}\left[\begin{array}{c} P \\ Q \end{array} \right] - \left[\begin{array}{c}\vec{p}_\mathrm{d} \\ \vec{q}_\mathrm{d} \end{array} \right] , \label{eq:mapping}
\end{equation}

\noindent where $\vec{p}_\mathrm{d},\vec{q}_\mathrm{d} \in \mathbb{R}^{n_b-1 \times 1} $ are the distribution grid's active and reactive power demands and $\vec{T}$ is the matrix that distributes the aggregated powers ($PQ$) among the individual power setpoints of $n_{g}$ generators in the distribution grid. $\vec{C}_\mathrm{g} \in \mathbb{R}^{n_b-1 \times n_g}$ is the generator to bus mapping matrix and the $GSK_{p,q} \in \mathbb{R}^{n_g \times 1}$ are the generation shift keys (GSKs) for the active and reactive generator powers. For example, the GSKs could weight the distribution among the generators according to their maximum generation capability ($\vec{p}_{\max},\vec{q}_{\max} \in \mathbb{R}^{n_g \times 1} $) which can be stated as
\begin{align}
GSK_p & =  \vec{p}_{\max} (\vec{1}^T\vec{p}_{\max})^{-1} \ , \\
GSK_q & =  \vec{q}_{\max} (\vec{1}^T\vec{q}_{\max})^{-1}  \ .
\end{align}

\subsection{Derivation of Polyhedral Sets}
A polytope is a convex polyhedral set that can be described with linear inequalities \cite{grünbaum2003convex}. Here, we express these linear constraints as a function of the active and reactive power (P,Q) flowing through one of the distribution grid feeders, such that the resulting PQ capability area can be shown as a polygon. Only the binding inequalities form the polygon. In this way the original number of constraints is reduced. Note that without any restriction this method can be expanded to more dimensions. This would be the case if more feeders of a distribution grid are connected to several nodes of the transmission grid. 

\subsubsection{Polyhedral Generator Constraints}
The generator constraints form the following rectangle
\begin{equation}
\left[\begin{array}{c} \vec{1}^T \vec{p}_{\min}  \\ \vec{1}^T \vec{q}_{\min}  \end{array} \right] \leq \left[\begin{array}{cc}  1 & 0 \\ 0 & 1\end{array} \right] \left[\begin{array}{c} P  \\ Q  \end{array} \right] \leq \left[\begin{array}{c} \vec{1}^T \vec{p}_{\max} \\ \vec{1}^T \vec{q}_{\max}  \end{array} \right] , \label{eq:polyg}
\end{equation}
\noindent where $\vec{p}_{\min},\vec{q}_{\min}$ are the minimum generation capabilities in terms of active and reactive power.

\subsubsection{Polyhedral Voltage Constraints}
The voltage magnitudes $\vec{v}$ should be in the range of the minimum and maximum voltage magnitudes $\vec{v}_{\min}, \vec{v}_{\max}$. This is achieved by inserting \eqref{eq:mapping} into \eqref{eq:vapprox} and bounding $\vec{v}$ within $\vec{v}_{\min}, \vec{v}_{\max}$, such that the associated constraint set is given as follows
\begin{align}
 VTDF \ \vec{T} \left[\begin{array}{l} P\\ Q \end{array}\right] & \leq \vec{1}v_{\max} - \vec{1}v_0 + VTDF \left[\begin{array}{l} \vec{p}_d \\ \vec{q}_d \end{array}\right], \nonumber \\
 -VTDF \ \vec{T} \left[\begin{array}{l} P\\ Q \end{array}\right]& \leq -\vec{1} v_{\min} + \vec{1}v_0 - VTDF \left[\begin{array}{l} \vec{p}_d \\ \vec{q}_d \end{array}\right] . \label{eq:polyv}
\end{align}

\subsubsection{Polyhedral Branch Flow Constraints}
As derived in \cite{Fortenbacher2019} we can approximate the circular operating area of the apparent line powers $\vec{s}$ with polygons. Here, their constructions are expressed in matrix form as
\begin{equation}
-\underbrace{\left[\begin{array}{c} \vec{s} \\ \vec{s} \\ \vec{s} \\ \vec{s}\end{array} \right]}_{\vec{S}} \leq \underbrace{\left[\begin{array}{cc} \vec{I} & \vec{A}_q \\  \vec{I} & -\vec{A}_q \\ \vec{A}_q & \vec{I} \\ \vec{A}_q & \vec{I} \end{array} \right]}_{\vec{B}} \left[\begin{array}{c}\vec{p}_\mathrm{f} \\ \vec{q}_\mathrm{f} \end{array} \right] \leq \underbrace{\left[\begin{array}{c} \vec{s} \\ \vec{s} \\ \vec{s} \\ \vec{s}\end{array} \right]}_{\vec{S}} , \label{eq:sapprox}
\end{equation}
where $\vec{A}_q$ defines the line segments of the polygon. By inserting \eqref{eq:branchapprox} into \eqref{eq:sapprox}, we can formulate the linear inequalities as follows
\begin{align}
\vec{B} \ PTDF \ \vec{T} \left[\begin{array}{l} P\\ Q \end{array}\right] & \leq \vec{S} + \vec{B} \ PTDF \left[\begin{array}{l} \vec{p}_d \\ \vec{q}_d \end{array}\right] , \nonumber \\
-\vec{B} \ PTDF \ \vec{T} \left[\begin{array}{l} P\\ Q \end{array}\right]& \leq \vec {S} - \vec{B} \ PTDF \left[\begin{array}{l} \vec{p}_d \\ \vec{q}_d \end{array}\right] . \label{eq:polybranch}
\end{align}

\section{Results}
We present the results for four different distribution grids ranging from low voltage to medium and high voltage networks. To compute and plot the polyhedral sets, we use the MPT toolbox \cite{MPT3}. For the validation, we compute a sequence of optimal power flow (OPF) problems by fixing the aggregated active power $P$ and gradually increasing it from $\vec{1}^T \vec{p}_{\min}$ to $\vec{1}^T \vec{p}_{\max}$, while fixing the active powers of the distribution grid units with the $GSK_p$. At every iteration, we formulate $Q$ as a decision variable and set a cost associated with $Q$ to utilize the full reactive power potential. In addition, we include linear constraints to the OPF problems that couple $Q$ with the reactive powers of the distribution grid units by using the $GSK_q$. We need to scan in two directions $+Q,-Q$. For the upper direction, we set a negative cost on $Q$, while for the lower one, we set a positive cost on $Q$. In total, we solve 200 OPF problems to track and compose the complete PQ capability area. 

\begin{figure}[t]
	\centering
	\subfloat[CIGRE LV network.]{\includegraphics[width=0.9\columnwidth]{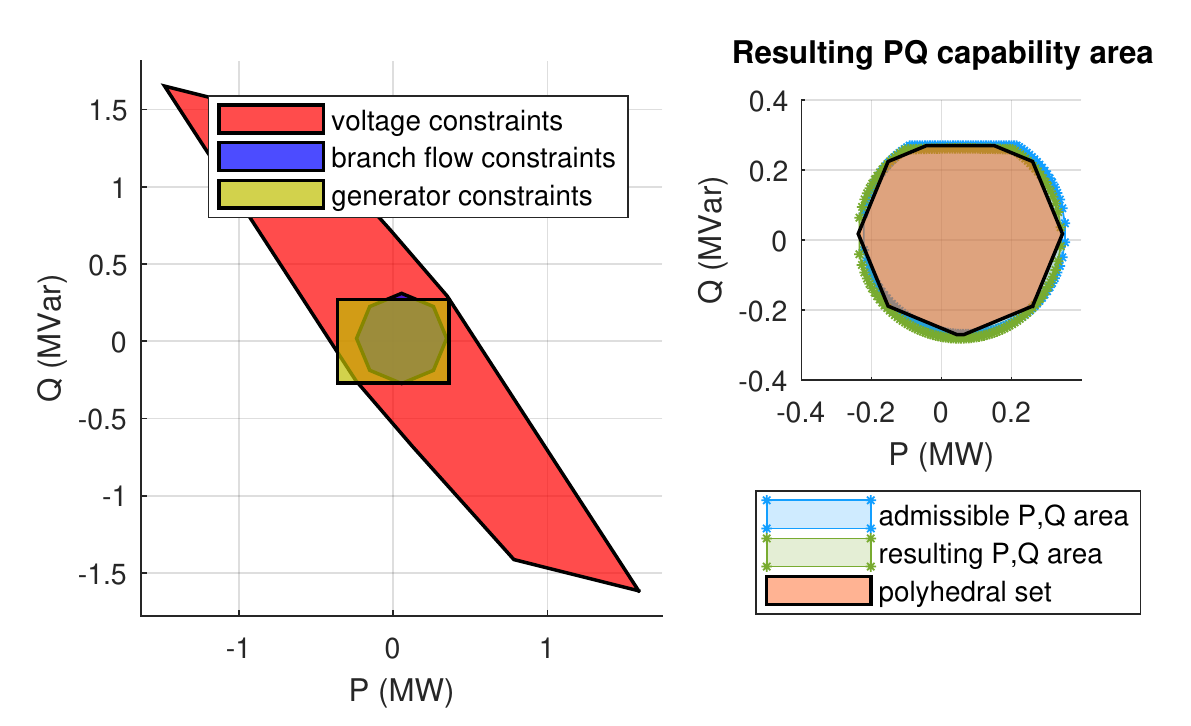}\label{fig:lv_cigre}}
	
	\subfloat[CIGRE MV network]{\includegraphics[width = 0.9\columnwidth]{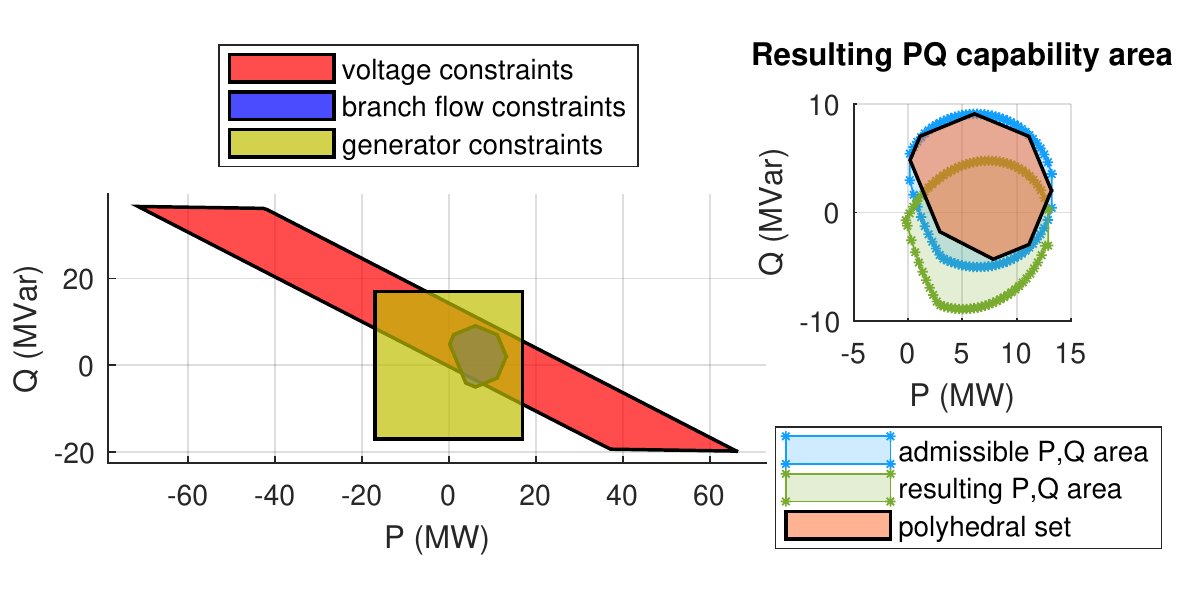}  
		\label{fig:cigreMV}}
	\caption{Composition of the polyhedral PQ capability area (left plots) and the resulting PQ capability area (right plots).}
	\label{fig:cigre}
\end{figure}

As an example of results, we calculate and plot PQ capability areas (Fig.~\ref{fig:cigre}) for the CIGRE LV grid (Fig.~\ref{fig:lv_cigre}) and the CIGRE MV grid (Fig.~\ref{fig:cigreMV}) under a given loading. On the left plots, the red polygon defines the operating region for the distribution grid feeder, in which the voltage of the distribution grid at all nodes stays between $v_{\min}$ and $v_{\max}$ and is represented by~\eqref{eq:polyv}. The deep blue region is associated with the branch flow constraints~\eqref{eq:polybranch} and is mainly limited by the transformer rating of the distribution grids. The light blue rectangle represents the generator constraints~\eqref{eq:polyg}. On the right plots, the resulting PQ area combining constraints~\eqref{eq:polyg},\eqref{eq:polyv}, and~\eqref{eq:polybranch} is shown labeled with polyhedral set. We show two areas obtained from the OPF solutions. The admissible PQ area defines the aggregated PQ setpoints of all distribution grid generators. It can be regarded as the generation potential that can be activated by the TSO without violating any distribution grid constraint. The resulting PQ area is the accessible power at the feeder covering the active and reactive power losses. The admissible PQ area in Fig.~\ref{fig:lv_cigre}  matches quite well with the resulting one, while in Fig.~\ref{fig:cigreMV} the areas are shifted in the $Q$ direction. This shift is mainly caused by the reactive power losses of the transformer. However, the admissible area, which is of more interest, still matches quite well with the polyhedron. 

\subsection{Accuracy}
To validate the accuracy of our approach, we introduce the following performance indicators. To define the approximation error, we intersect the admissible PQ area ($A_\text{admiss}$) and the polyhedral PQ capability area ($A_\text{poly}$). The error is defined as follows: 
\begin{equation}
\text{Error} = 1-  \frac{A_{\text{admiss}}\cap A_{\text{poly}}}{A_{\text{poly}}}
\end{equation}

In addition, we define the fill factor, which is a measure of which extent the polyhedral area covers the admissible area, defined as follows: 
\begin{equation}
\text{Fill Factor} = \frac{A_{\text{admiss}}\cap A_{\text{poly}}}{A_{\text{admiss}}} .
\end{equation}
\begin{table}[t]
	\centering
	\caption{Results for different distribution grids.}
	\begin{tabular}{lrrrrr}
		\hline
		Grid & Buses   & Error & Fill Factor & \multicolumn{2}{c}{Computing Time (sec)}\\
		& (-)     & (\%) & (\%) & polyhedral & OPF \\
		\hline 
		CIGRE LV \cite{cigre} & 19 & 0.47  & 92.2 & 0.07 & 9.32 \\
		CIGRE MV \cite{cigre}& 18 & 0.04 & 81.2 & 0.08 & 7.59\\
		oberrhein \cite{pandapower}& 184& 9.2 & 80.3 & 0.14 & 61.8\\
		utlility for Zurich  & 407& 0 & 38.2 & 0.13 & 74.4\\
		\hline
	\end{tabular}
	\label{tab:results}
\end{table}
Table~\ref{tab:results} lists the results on the accuracy. We find that the approximation errors are low except for the oberrhein grid. The higher error can be explained that this grid has high branch charging susceptances that cannot be compensated with the reactive power losses in our approach. Thus, this leads to a small shift of the polyhedral set in postive Q direction. The fill factors are high except for the grid from the utility for Zurich. It can be anticipated that for this HV grid, the linear lossless approximation is not as good as for distribution grids, since the errors on the voltage magnitudes increase due to higher reactive power losses caused by a higher X/R ratio.

\subsection{Complexity}
\begin{figure}[t]
	\centering
	\includegraphics[width=0.9\columnwidth]{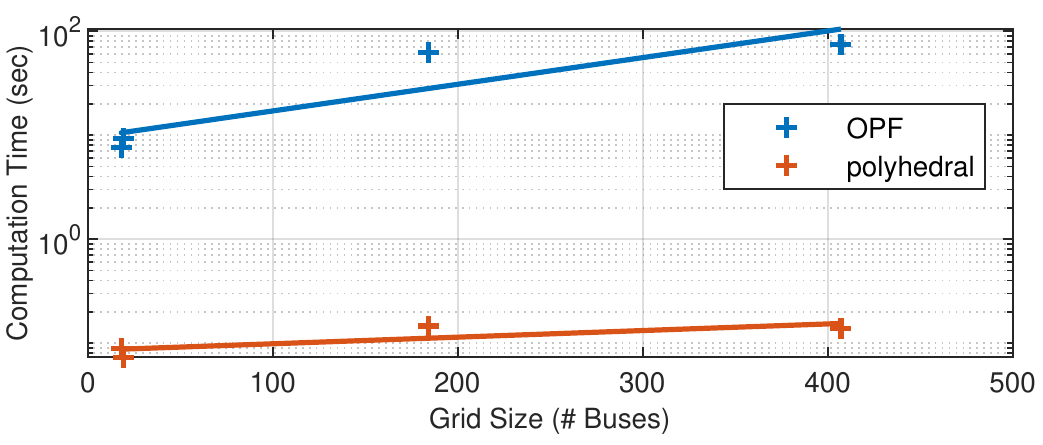} 
	\caption{Complexity comparison.} \label{fig:complexity}
\end{figure}
The plot in Fig.~\ref{fig:complexity} shows the computational times listed in Table~\ref{tab:results} in semi-logarithmic scale as a function of the number of buses for computing the polyhedral sets and the OPF solutions. We also compute their semi-logarithmic regressions. The slope of the polyhedral approach is much lower than of the OPF approach and is faster by a factor of 500 at 400 buses. Moreover, the OPF approach would be clearly intractable for higher dimensions, since the number of OPF calculations explodes exponentially with the number of distribution grid feeders.

\section{Conclusion}
In this paper we have presented a novel tractable method to compute a reduced and aggregated representation of distribution grids. We derive polyhedral sets based on an existing linear lossless power flow approximation to obtain a PQ capability area, in which all distribution grid constraints are satisfied. The PQ capability area is accessible at the distribution grid feeder, representing the interface between the TSO and DSO, and denotes the available flexibility for the TSO. Since we use a lossless power flow approximation, we observe a small shift of the approximated PQ area in the active power $P$ direction to the resulting PQ area, while for MV/HV networks this shift is more pronounced in the reactive power $Q$ direction. This can be explained by the fact that in MV/HV networks the reactive power losses are predominant and cannot be captured by the approximation. However, the admissible PQ areas match well for all grids and these are of more interest for the TSO, since those reflect the generation potential that can be activated from the distribution grid. Future work will focus on extending this work by including more distribution grid feeders in the transmission network and transformer tap ratio or phase shifting control. This would introduce more dimensions and require the computation of polytopes.    

\section*{Acknowledgment}
This research is part of the activities of T\&DFlex -- TSO-DSO Flexibility: towards integrated grid control and coordination in Switzerland, which is financially supported by the Swiss Federal Office of Energy (SFOE) and the Swiss Association for Energy and Network Research (SGEN).

\bibliography{literature}

\end{document}